\documentclass[preprint,3p,12pt,twocolumn]{elsarticle}
\usepackage[english]{babel}

\usepackage{siunitx}
\usepackage{subcaption}
\let\qty\SI

\usepackage{booktabs}
\usepackage{multirow}

\usepackage{amssymb}
\usepackage{amsmath}

\usepackage{lineno}

\journal{Nuclear Instruments and Methods in Physics Research Section A}

\begin{document}
	
	\begin{frontmatter}
		

		\title{Implementation of a high-efficiency muon veto system for the GeSparK alpha-beta/gamma coincidence detector}
		\author[1,2]{A.~Barresi\corref{cor}}
		\cortext[cor]{Corresponding authors: \newline andrea.barresi@unimib.it, \newline massimiliano.nastasi@unimib.it}
		\author[1,2]{D.~Chiesa}
		\author[1,2]{M.~Nastasi\corref{cor}}
		\author[1,2]{E.~Previtali}
		\author[1,2]{M.~Sisti}
		\address[1]{Department of Physics, University of Milano-Bicocca, 20126 Milan, Italy}
		\address[2]{INFN, Milano-Bicocca, 20126 Milan, Italy}
		
		\begin{abstract}
			
			One of the main background sources of low background high purity germanium (HPGe) detectors is the cosmic muon showers produced in the interaction with the lead and copper shield surrounding the detector that can induce signals not distinguishable from radioactivity events.
            Plastic scintillators are widely used to implement active veto systems to reduce this contribution and increase the measurement sensitivity.
            In this work, we present the implementation of a high-efficiency muon veto system for the alpha-beta/gamma coincidence detector, called GeSparK. The veto system consists of six plastic scintillator detectors accurately positioned around the HPGe and liquid scintillator detectors. The final design includes two detectors above and four inside the passive shielding between the copper and lead layers. In this way, we could limit both their size and the event rate produced by external radioactivity. A dedicated background measurement showed that the achieved background reduction is a bit less than 93\%.
			
		\end{abstract}
		
		\begin{keyword}
			
			Low background gamma-ray spectroscopy \sep Coincidence measurements \sep Muon veto \sep Liquid scintillator \sep HPGe detector \sep Plastic scintillator
			
		\end{keyword}
		
	\end{frontmatter}
	
	
	
	\section{Introduction}
	\label{S:1}

\begin{figure}[]
		\begin{center}
			\includegraphics[width=0.45\textwidth]{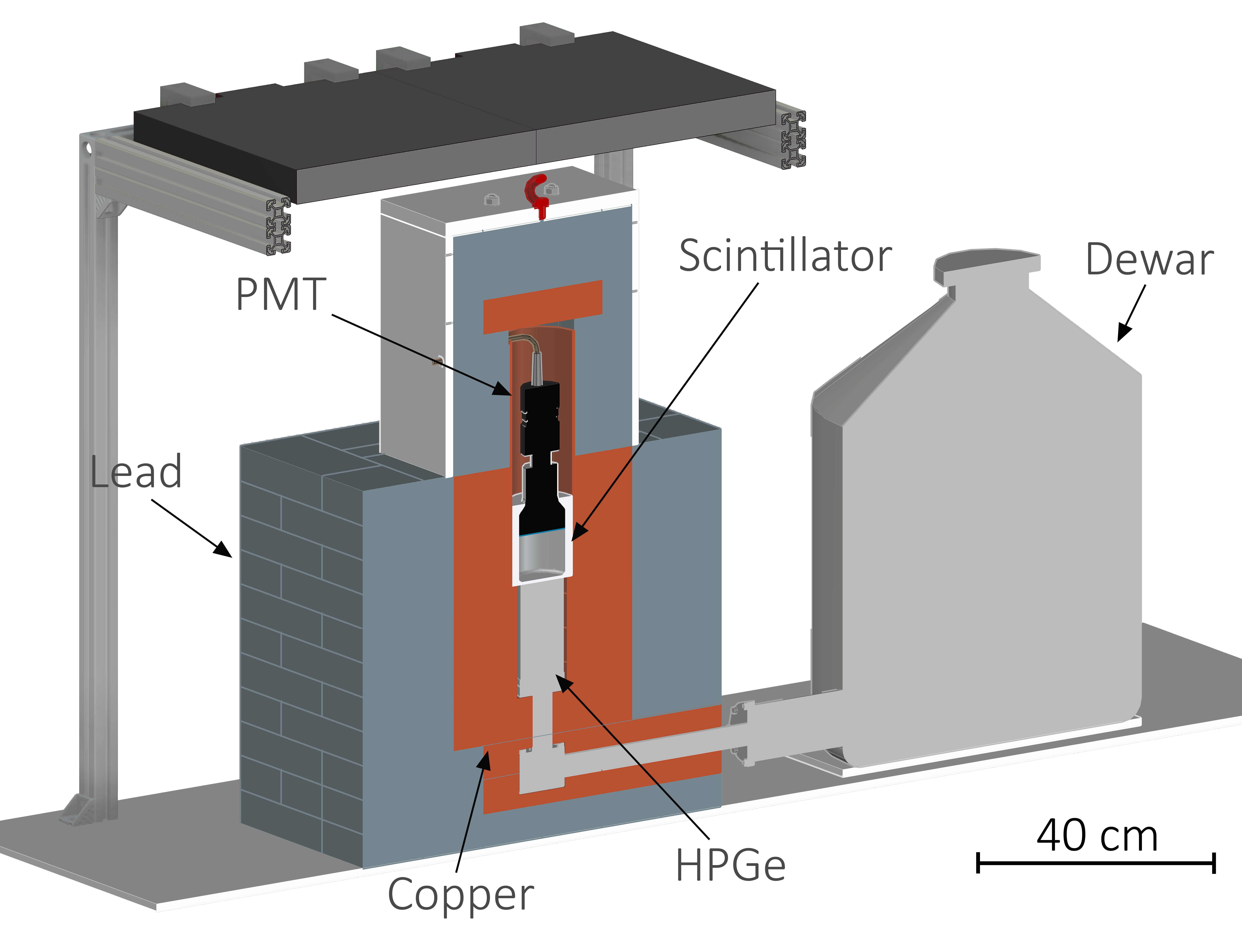}
			\caption{Section of the GeSparK detector with passive shielding and top muon veto system.}
			\label{fig:GeSparKsection}
		\end{center}
	\end{figure}

    The most recent rare events physics experiments require increasingly lower levels of radiopurity for their materials, therefore screening techniques also require the development of new detectors and techniques to reach the needed sensitivities. 
	In a previous paper \cite{GS:2021} we described a new low background detector, GeSparK, developed at the Radioactivity Laboratory in the Department of Physics of the University of Milano-Bicocca. It is a composite system of a Liquid Scintillator (LS) and a High Purity Germanium (HPGe) detectors, working in time coincidence mode to allow for the acquisition of decay events characterized by well-defined time correlations, such as decay cascades or alpha-gamma and beta-gamma coincident events, thus drastically reducing the environmental and cosmogenic backgrounds (Figure \ref{fig:GeSparKsection}).
    Although the coincidence between the LS and HPGe detectors provides a strong background reduction, the residual event rate is strongly dominated by cosmic muons, whose showers can produce coincident events characterized by a continuous energy spectrum. A common approach to reduce this background source is the implementation of active shieldings to veto the cosmic muons by means of plastic scintillator detectors usually installed above the passive shielding, as originally implemented in the GeSparK detector.
    The limitation of this approach is the low angular coverage of the veto system which limits the veto efficiency.
	In this work, we describe the implementation of an improved muon veto system for the GeSparK detector composed of six plastic scintillator detectors accurately positioned around the HPGe and LS detectors: two above and four inside the passive shielding between copper and lead layers. This strategy allowed us to use small-size plastic scintillators, compared to those required by an external installation, thus reducing the cost of the system without affecting the final veto efficiency.

	\section{The GeSparK detector structure}
	\label{S:2}

	The GeSparK system comprises HPGe and LS detectors with the following features.
	
	The HPGe detector is an Ortec coaxial p-type with a relative efficiency of 38\% and an energy resolution of \qty{1.8}{keV} at \qty{1332}{\keV}. The detector end-cap is equipped with a thin carbon window to increase the detection efficiency of low-energy gamma-ray and X-ray photons.
	
	The scintillator detector consists of a cylindrical PTFE (polytetrafluoroethylene) container with a thin bottom thickness (\qty{2}{mm}), to minimize the absorption of low-energy photons and a cap of PMMA (polymethylmethacrylate) which, being transparent, allows coupling with a low-background \qty{3}{inch} PMT by Hamamatsu.
	The volume of the container is \qty{200}{\mL} and it is filled with the sample to be screened mixed to the liquid scintillator (Ultima Gold AB - Perkin Elmer \cite{Edler:2016}) up to 1:1 ratio. 
		
	Both detectors are surrounded by copper and lead shields to reduce the influence of environmental radioactivity, as shown in Figure \ref{fig:GeSparKsection}. On the bottom of the structure, around the HPGe detector and the sample container, there are \qty{10}{cm} of copper surrounded by \qty{15}{cm} of lead; the shield thicknesses are reduced to \qty{5}{mm} of copper and \qty{10}{cm} of lead around the PMT. Finally, \qty{5}{cm} of copper and \qty{10}{cm} of lead are placed on top of the structure. To reduce the background contribution produced by the secondary radiation associated with cosmic muons, two plastic scintillators $(80\:\textrm{cm}\times 40\:\textrm{cm}\times5\:\textrm{cm})$ were originally installed on top of the shield to perform an anti-coincidence selection for the signal recorded by both detectors. With this configuration, a reduction efficiency of \qty{53.2(3)}{\%} of the total background rate was achieved \cite{GS:2021}.
	
    The acquisition system was updated with respect to that described in the previous paper \cite{GS:2021}. In particular, the output signal of the PMT is split and acquired in parallel by two channels with different dynamic ranges of the NI oscilloscope mod.\ PXI 5153 with a sampling frequency of \qty{1}{GHz} and a resolution of 8 bits.
	The simultaneous acquisition of the PMT signal with two dynamic ranges allows us to acquire both low-energy and high-energy alpha and beta signals making the most of the limited resolution of the fast board.

    \section{Monte Carlo simulations and dedicated measurements for muon veto efficiency evaluation}
	\label{S:3}

    To increase the sensitivity of the GeSparK detector we planned to implement a new muon veto system able to reach an almost unitary veto efficiency. 
    Since this implementation would have required a non-negligible effort to redesign and modify the previous shielding, we decided to exploit Monte Carlo simulations of the complete GeSparK system to evaluate the expected veto efficiencies for different configurations and determine the best one.
	
	The preliminary characterization of the muon flux in the laboratory, the MC validation, and the final configuration efficiency are discussed in the next sections, followed by the result of the veto efficiency measurement after the veto system implementation.
    
	\subsection{Muon flux characterization}
	\label{S:3.1}
 
	The radioactivity laboratory of the physics department of the University of Milano-Bicocca is located at the third underground floor of the department building. This location could modify the angular distribution of the muon flux due to different absorption at different angles as a consequence of the asymmetric structure of the building.

\begin{figure}
		\begin{center}
			\includegraphics[width=0.45\textwidth]{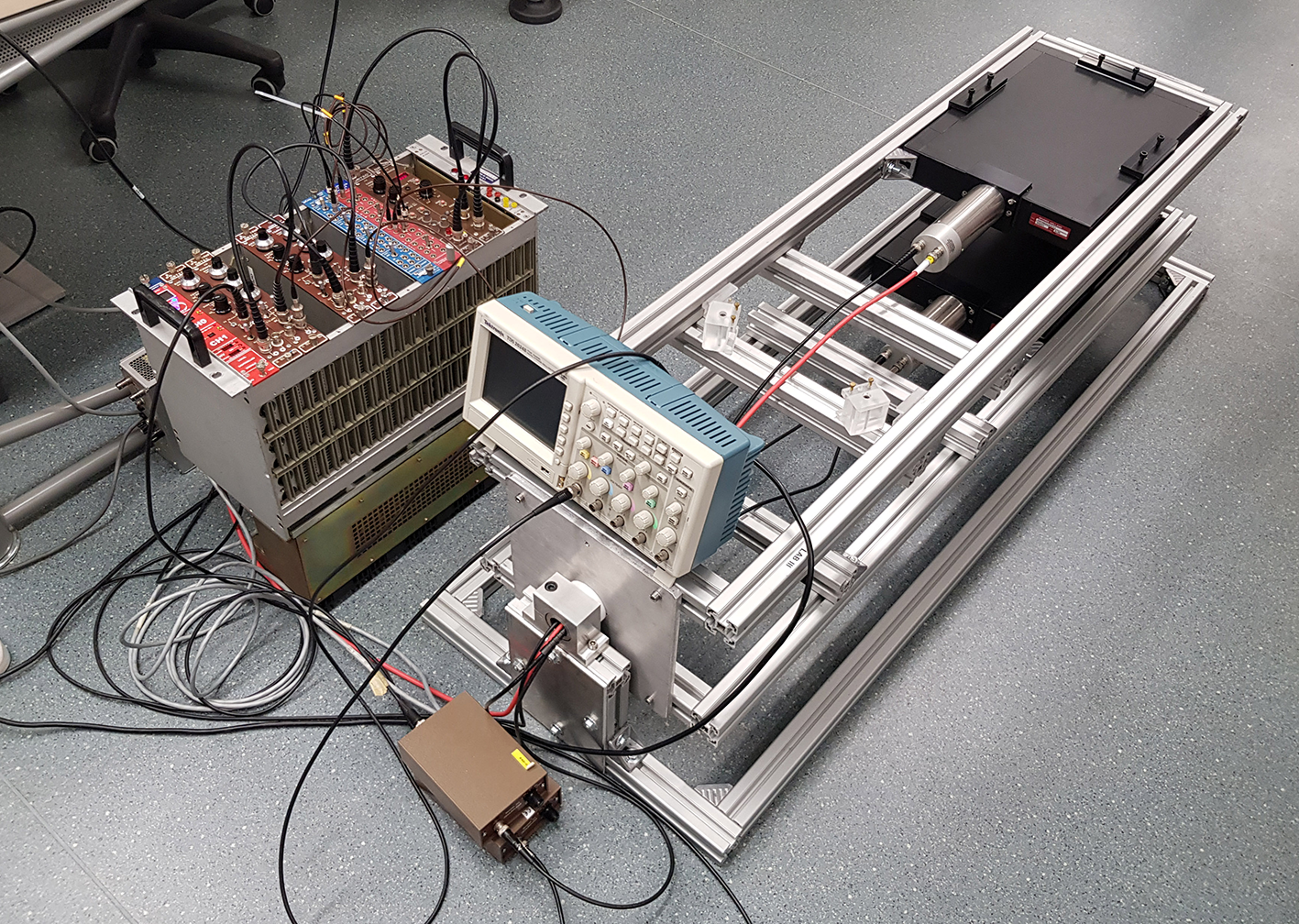}
			\caption{Experimental setup used for the muon flux characterization. It consists of two plastic scintillators \qtyproduct{300x420x50}{\mm}, each installed on a rotating support at a fixed distance of \qty{30}{cm}, working in time coincidence by exploiting a NIM module chain.}
			\label{fig:MuonExpSetup}
		\end{center}
	\end{figure}

\begin{figure}[t]
		\begin{center}
			\includegraphics[width=0.45\textwidth]{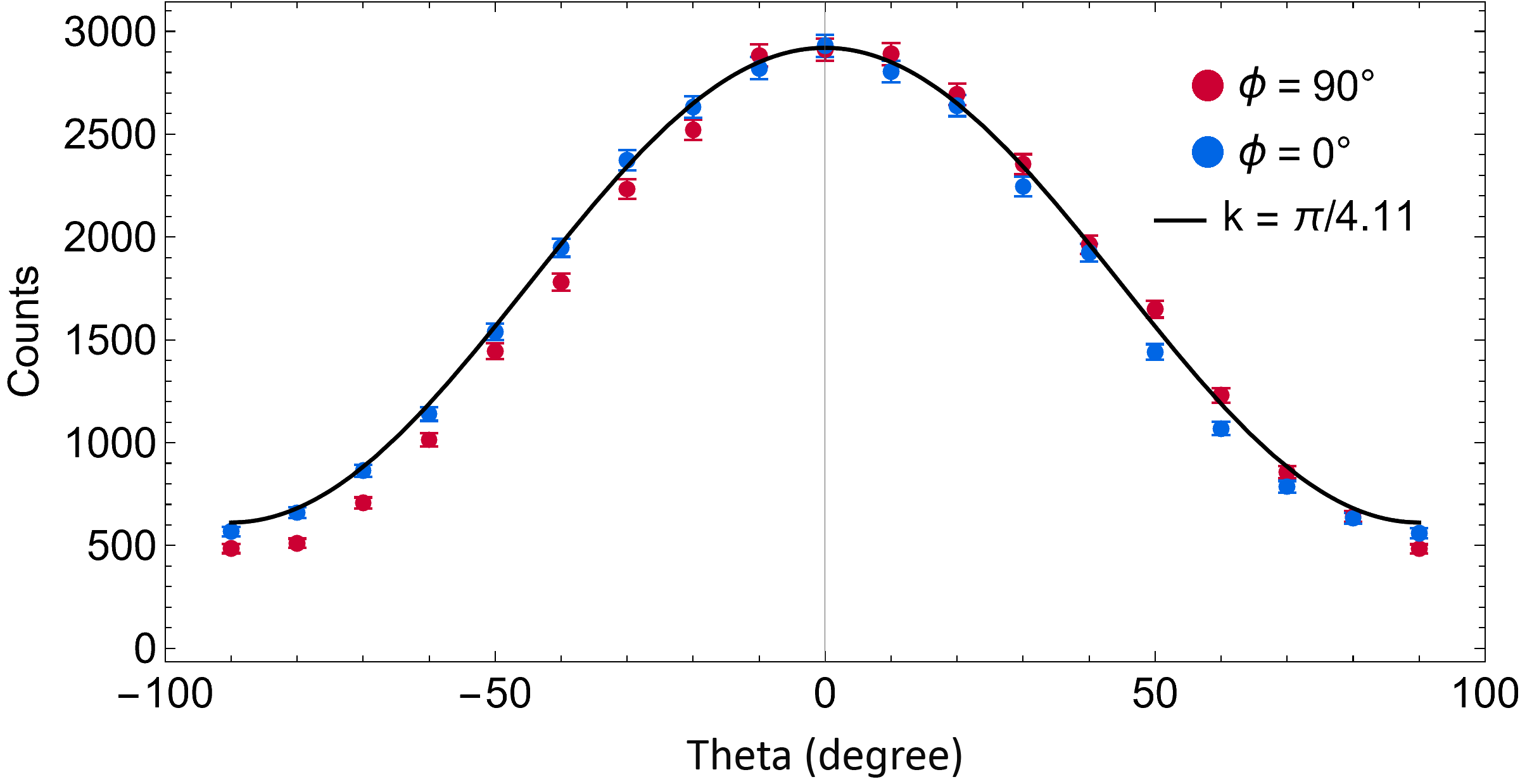}
			\caption{Measured muon angular distribution in the radioactivity laboratory on the third underground floor of the Physics department of the University of Milano-Bicocca. The data are fitted with equation \ref{eq:MuAngDist}}
			\label{fig:MuonAngularDistributions}
		\end{center}
	\end{figure}

	To understand this effect, dedicated measurements of the muon angular distribution were performed along two orthogonal planes, parallel and perpendicular to the department building. The used muon telescope consists of two plastic scintillators \qtyproduct{300x420x50}{\mm}, each installed on a rotating support at a fixed distance of \qty{30}{cm}, as shown in Figure \ref{fig:MuonExpSetup}, and working in time coincidence by exploiting a NIM module chain. For each plane, the integral muon flux was measured every \qty{10}{\degree} from \qty{-90}{\degree} to \qty{+90}{\degree} ($\theta$) at two different orientations ($\phi$), parallel (\qty{0}{\degree}) and perpendicular (\qty{90}{\degree}) to the building structure.
	This measurement showed that the overall angular distribution of muons in the radioactivity laboratory follows with good agreement a $ \cos^2(\theta) $ distribution with small deviations probably due to asymmetrical absorption in the soil and buildings, as shown in Figure \ref{fig:MuonAngularDistributions}. In the plot it is also shown the best fit based on a simple model that was constructed to interpret the experimental results and verify deviations from the perfect $ \cos^2(\theta) $ distribution. The experimental setup is not able to measure the exact angular distribution of muons because a muon that can induce a signal in both detectors can come from a range of angles. In other words, the measurement system integrates the muon angular distribution in an angular range given by the geometrical properties of the detectors. The integral goes from $ \theta-k $ to $ \theta+k $, where $\theta$ is the angle formed by the axis perpendicular to the plastic scintillator with the vertical axis and k is the semi-range angle of the field of view of the system.
	Since the distance of the two plastic scintillators is about equal to the shortest side of the detectors, the k parameter is expected to be around $ \pi/4 $.
	The function that describes the data can be computed in the following way:
	\begin{equation}
    \begin{aligned}
		I(\theta) &= a\cdot\frac{\int_{\theta-k}^{\theta+k}\cos^2(\alpha)d\alpha}{\int_{-k}^{k}\cos^2(\alpha)d\alpha} \\ &= a \cdot \frac{\cos (2 \theta ) \sin (k) \cos (k)+k}{k+\sin (k) \cos (k)}
		\label{eq:MuAngDist}
    \end{aligned}
	\end{equation}
	where "a" is a coefficient that represents the number of counts at $ \theta = \qty{0}{\degree} $ and the denominator is simply a normalization factor to obtain that the maximum of the function is equal to "a".

	To have an accurate estimation of the k value with respect to the simple consideration described before that can be used to describe the experimental data, a simulation was performed.
	The simulations were performed for each angular position (every \qty{10}{\degree} from \qty{-90}{\degree} to \qty{90}{\degree}) by simulating \num{100000} muons distributed as $ \cos^2(\theta) $ for each position and counting the number of events that produces a coincidence signal in the two detectors (see next session for more simulation details).
    The best fit of the data with the equation \ref{eq:MuAngDist} provided the best estimation for $ k = \pi/4.11 $ or \qty{43.8}{\degree}. This value is used to fix the k value in equation \ref{eq:MuAngDist} used to describe the experimental data with "a" fixed to the data value at $ \theta = \qty{0}{\degree} $.
	
    In conclusion, since the observed differences due to asymmetric absorptions are quite small they were not directly considered in the simulations performed for the GeSparK detector.

    \subsection{Monte Carlo simulations validation}
	\label{S:3.2}

	A Monte Carlo simulation of the whole structure of the GeSparK detector was implemented to determine the response of the plastic scintillator detectors and their veto efficiency.
	These simulations were performed by using a simulation software based on Geant4 \cite{Agostinelli:2003}, called Genesis, developed at the Department of Physics of the University of Milano-Bicocca.
	To perform these simulations with the best accuracy, the following information must be provided to Genesis:
	\begin{itemize}
		\item Geometric model of GeSparK detector with plastic scintillators
		\item Angular distribution of the muon flux
		\item Energy distribution of the muons
	\end{itemize}
 
	The geometric model of the GeSparK detector is constructed by combining the CAD (computer-aided design) model developed during the construction of the detector and a detailed text-based description of the HPGe and scintillator detectors. The CAD model includes the geometry of the copper and lead shieldings, the liquid nitrogen dewar, and the cold finger of the HPGe. Using CAD, models of the support frame for the plastic scintillators and the plastic scintillators themselves were also created. Text geometry is instead used for the description of the HPGe detector, which comprises the germanium crystal, the holder, the end-cap, and the cold finger, as well as for the description of the LS detector.
 
	The Genesis software utilizes the General Particle Source (GPS) class of Geant4 to generate the primary particles. This class is versatile and allows defining a wide range of sources. In particular, the GPS allows the user to define the angular and energetic distributions of the primary particle, in this case, muons, by providing the histograms of the distributions.

	The energy distribution of the muons should not be a critical parameter of the simulation because the behavior of high-energy muons is not very different with respect to lower-energy ones since they are all around the minimum ionization region. However, to use a quite realistic distribution, instead of a monochromatic source, the energy histogram is obtained by approximating the distribution reported in \cite{MuoniEne} from the muons at \qty{0}{\degree} zenith angle at sea level from \qty{1}{\GeV} to \qty{200}{\GeV}.
	
	As far as the angular distribution is concerned we first used a $ \cos^2 $ distribution in $ \theta $ and a uniform distribution in $ \phi $. Since the GPS samples the $ \theta $ and $ \phi $ coordinates independently, to allow the GPS to generate the desired distribution, the provided histogram must include the $ \sin(\theta) $ term of the differential in the spherical coordinates system. In this case, the resulting distribution is given by $ \sin(\theta)\cdot\cos^2(\theta) $. A dedicated algorithm was developed in Genesis to generate events distributed on a semi-spherical surface with a generic radius (\qty{1.5}{\m} for these simulations) from the LS and HPGe detectors of GeSparK, with the angular distribution provided by the GPS class.

	The angular distribution of atmospheric muons is not constant but depends on the energy of the muons considered \cite{MuoniAng}. At the sea level if we integrate all the energetic contributions we obtain the usual $ \cos^2(\theta) $ distribution which is characteristic of the muons of about \qty{3}{\GeV} of energy. At lower energies, the angular distribution becomes increasingly steep, while at higher energies it flattens.
	Considering that the radioactivity laboratory is located at the third underground floor of the Department building, we can assume that a fraction of the kinetic energy of the muons is lost in the passage through the walls of the building, losing energy in the order of few GeV. 
    In this case, for a given particular energy, the angular distribution of the muons reaching the lab is not the same as the one at sea level for muons of the same energy but corresponds to the angular distribution of slightly more energetic muons.

\begin{figure}
		\begin{center}
			\includegraphics[width=0.45\textwidth]{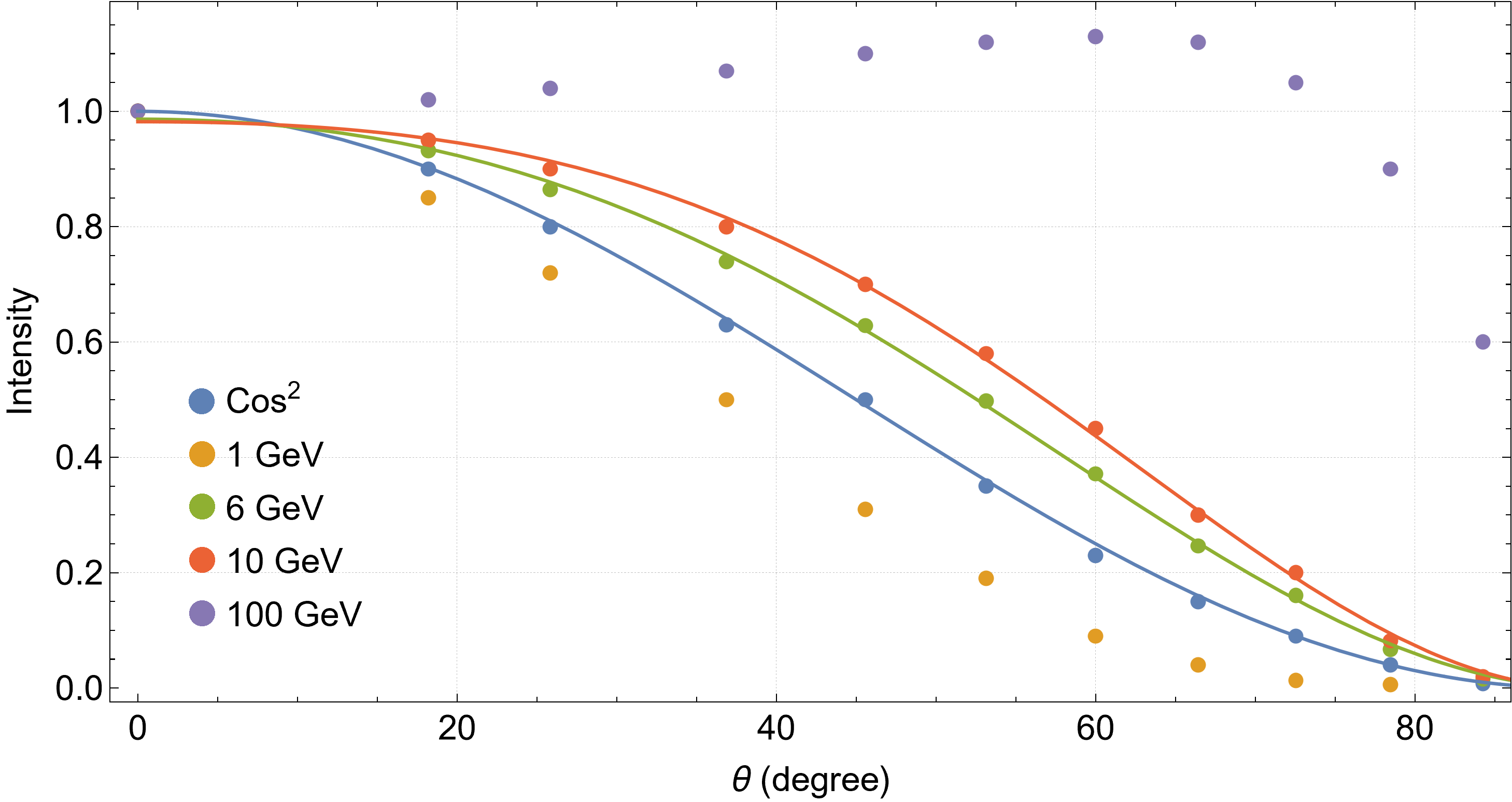}
			\caption{Angular distribution for atmospheric muons at different energies from the data reported in \cite{MuoniAng} (\qtylist{1;3;10;100}{\GeV}).}
			\label{fig:MCMuAngularDistrution}
		\end{center}
	\end{figure}

    To evaluate this effect, based on the data in reference \cite{MuoniAng}, the angular distribution of muons at different energies was obtained and used to generate the relative histogram distribution necessary for the Monte Carlo simulations.
	In Figure \ref{fig:MCMuAngularDistrution} the distributions for different energies are shown, from \qty{1}{\GeV} up to \qty{100}{\GeV}, and the increase of the contribution at large angles for high energy muons is visible.
	The data at \qtylist{1;3;10;100}{\GeV} were obtained directly from the plot in the reference \cite{MuoniAng}, while the distribution at \qty{6}{\GeV} was estimated by using a weighted mean between the \qty{3}{\GeV} and \qty{10}{\GeV} data. The PDF of the distribution at \qty{6}{\GeV} and \qty{10}{\GeV} were computed by fitting the data points with a parametric function. These functions were then multiplied for $ \sin(\theta) $ and normalized to obtain the PDFs. These PDFs were used to construct the histograms of the distributions required by Genesis. 

\begin{figure}[t]
		\begin{center}
			\includegraphics[width=0.45\textwidth]{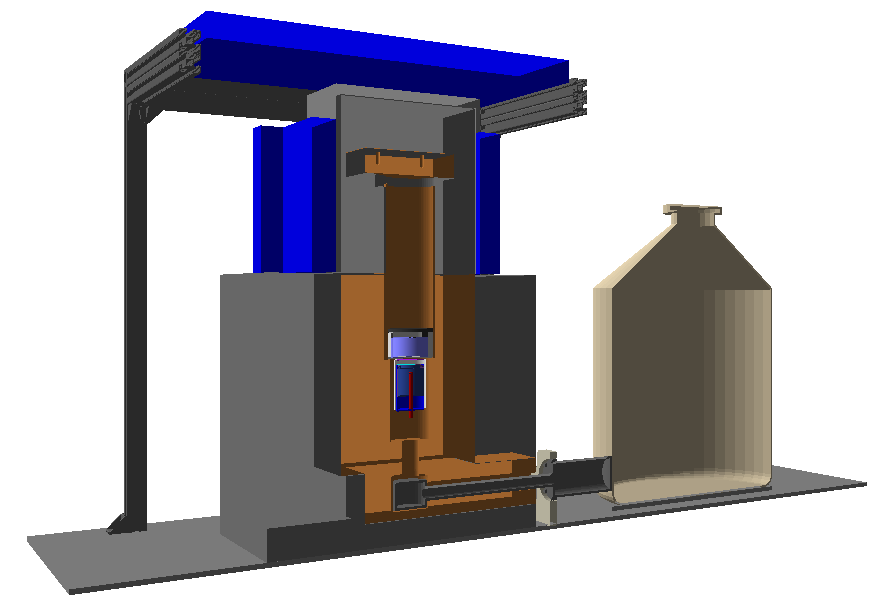}
			\caption{Complete Monte Carlo model of the GeSparK detector, including the HPGe and LS detectors, the passive shieldings, and the test muon veto system with external lateral plastic scintillators (blue).}
			\label{fig:GSMCModelOld}
		\end{center}
	\end{figure}

    To optimize and validate the Monte Carlo simulations with measured data, a test configuration of the veto system was implemented, consisting of the two original plastic scintillators \qtyproduct{1000x500x50}{\mm}, placed on the top, and four plastic scintillators \qtyproduct{300x420x50}{\mm}, placed on the side of the higher part of the lead shielding.
    The veto efficiencies were evaluated for four different combinations of plastic scintillators. The configurations are referred to as "Top," "2 Side," "Top + 2 Side," and "Top + 4 Side," indicating the specific plastic scintillators used for vetoing in each configuration.
    The Monte Carlo model of the test configuration is shown in Figure \ref{fig:GSMCModelOld}.
    
    \begin{table*}
		\centering
        \footnotesize
		\begin{tabular}{ccccc}
			\toprule
			Detectors & \multicolumn{4}{c}{Efficiencies} \\
			& \qty{3}{GeV} ($\cos^2$) & \qty{6}{GeV} &\qty{10}{GeV} & Measured \\
			\midrule
			Top & \qty{59.3(4)}{\%} & \qty{53.0(4)}{\%} & \qty{51.5(4)}{\%} & \qty{53.2(3)}{\%} \\
			2 side & \qty{33.1(4)}{\%} & \qty{32.7(4)}{\%} & \qty{33.6(4)}{\%} & \qty{36.3(5)}{\%} \\
			Top + 2 side & \qty{74.6(4)}{\%} & \qty{69.0(4)}{\%} & \qty{68.2(4)}{\%} & \qty{70.7(3)}{\%} \\
			Top + 4 side & \qty{87.0(3)}{\%} & \qty{82.6(3)}{\%} & \qty{81.8(3)}{\%} & \qty{81.7(1)}{\%} \\
			\bottomrule
		\end{tabular}
		\caption{Monte Carlo estimations for the plastic scintillators veto efficiency for different angular distributions compared to the experimental values. The reported errors include only the statistical contribution.}
		\label{tab:MuonVetoEfficiencyMCMIS}
	\end{table*}
    
\begin{figure}
		\begin{center}
			\includegraphics[width=0.45\textwidth]{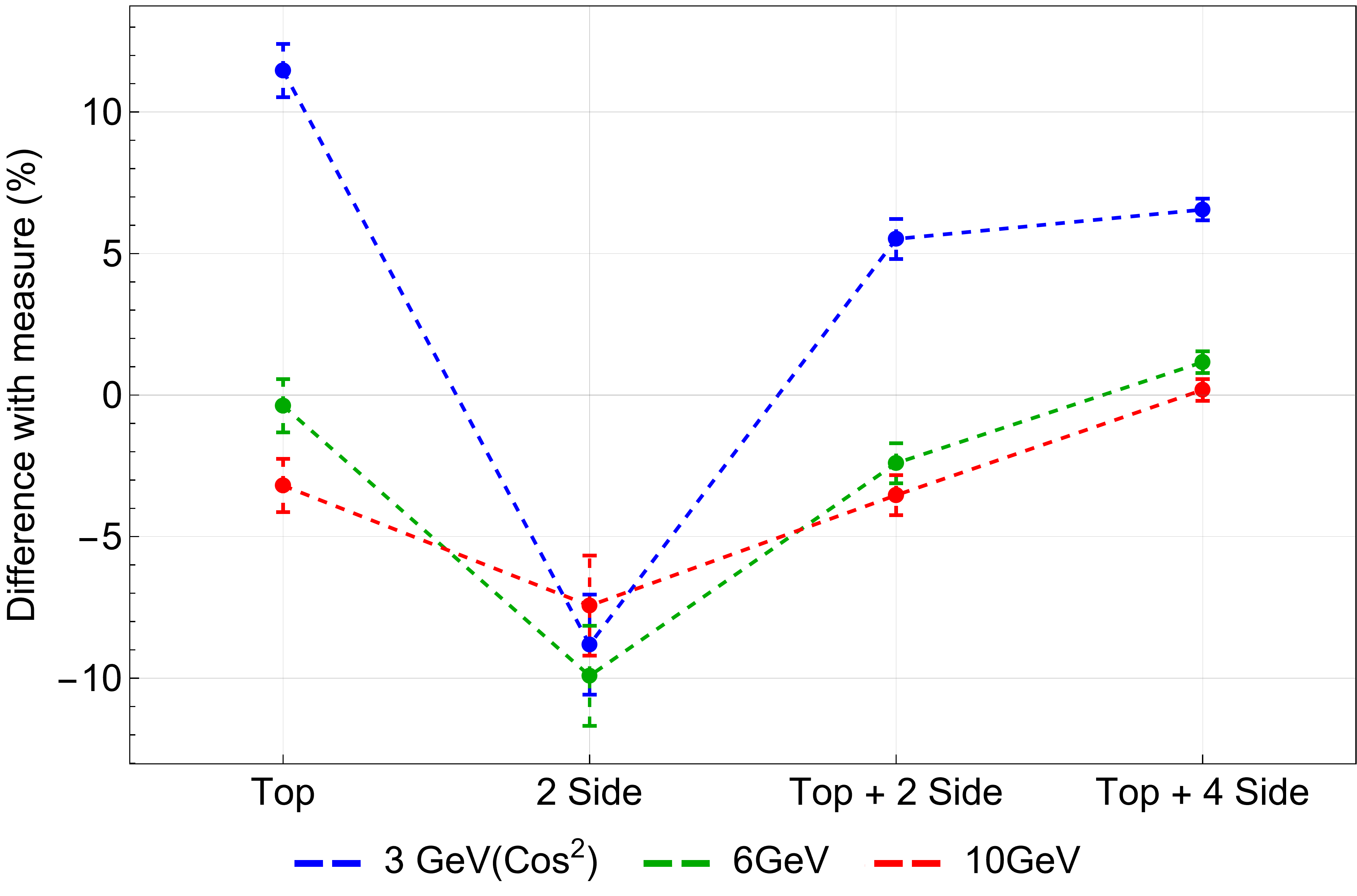}
			\caption{Percentage difference between simulated and measured veto efficiencies for the four plastic detector configurations and three angular distributions.}
			\label{fig:PercentageDifference}
		\end{center}
	\end{figure}

    In the Monte Carlo simulation, the detector selection is performed offline during data analysis by selecting only the data in coincidence with the required plastic scintillators. For the experimental measurements, four separate measurements were taken, one for each combination, by acquiring data only from the required plastic scintillators. The efficiency is calculated as the ratio between the number of events in coincidence with a signal in the plastic scintillators and the total number of events. The reported errors only include the statistical contribution based on the number of events. The results with different angular distributions are reported in Table \ref{tab:MuonVetoEfficiencyMCMIS} and the percentage differences between the simulated and measured efficiencies are reported in Figure \ref{fig:PercentageDifference}.
	
	The results at \qty{6}{\GeV} provide the most accurate estimation for the top detectors and a good agreement for the other configurations with an error much lower than the \qty{3}{GeV} one ($ \cos^2(\theta) $ distribution). This corroborates the hypothesis that the angular distribution is distorted with respect to the $ \cos^2(\theta) $ distribution due to the muon's kinetic energy loss and absorption in the materials. This condition can be used for further studies and improvement of the GeSparK veto system or other detectors in the radioactivity laboratory of the University of Milano-Bicocca.

    \begin{figure}
		\begin{center}
			\includegraphics[width=0.45\textwidth]{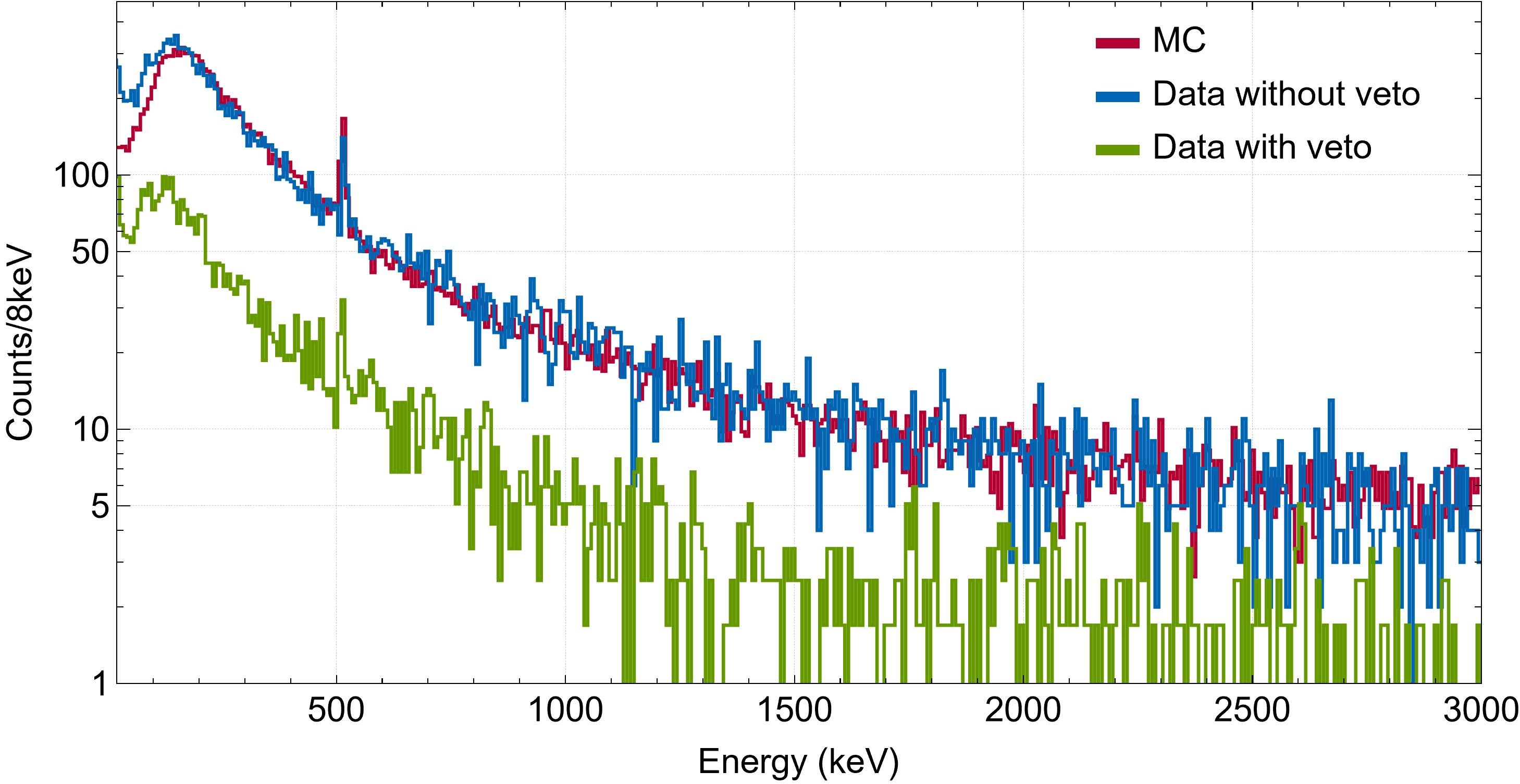}
			\caption{HPGe background spectra of the GeSparK detector with and without the application of the cosmic muon veto, obtained by measuring the coincidence events between LS and HPGe detector in the absence of radioactive sources in the LS. In the plot is superimposed the simulated energy deposition in the HPGe detector produced by cosmic muons, normalized for the total number of events in the measured spectrum without veto.}
			\label{fig:MCHPGeMuonSpectra}
		\end{center}
	\end{figure}
	The data produced by the simulation can also be compared to the residual background of the GeSparK detector after the application of the coincidence event selection. In this case, the detected events are associated only with signals produced in coincidence in both the LS and HPGe detectors.
	In Figure \ref{fig:MCHPGeMuonSpectra} the spectra of the residual background of the GeSparK detector are shown with and without the application of the cosmic muon veto, obtained by acquiring the coincidence events in the absence of a sample in the LS detector. In the same plot, the distribution of the energy deposited in the HPGe detector obtained by the simulation and normalized for the number of events in the measured spectrum is superimposed. The simulation reproduces with very good agreement the shape of the GeSparK background in the whole spectrum except for the very low energy region. This agreement and the absence of any particular peak in the background spectrum, except the \qty{511}{\keV}, which is produced by the annihilation of positrons produced by the muon showers, indicates that the background of the GeSparK detector is dominated by the contribution of the cosmic muon showers that can be produced when they interact with the passive shielding of the GeSparK system.

	\subsection{The new veto system and expected efficiency}
	\label{S:3.3}
 
	\begin{figure}[t]
		\begin{center}
			\includegraphics[width=0.45\textwidth]{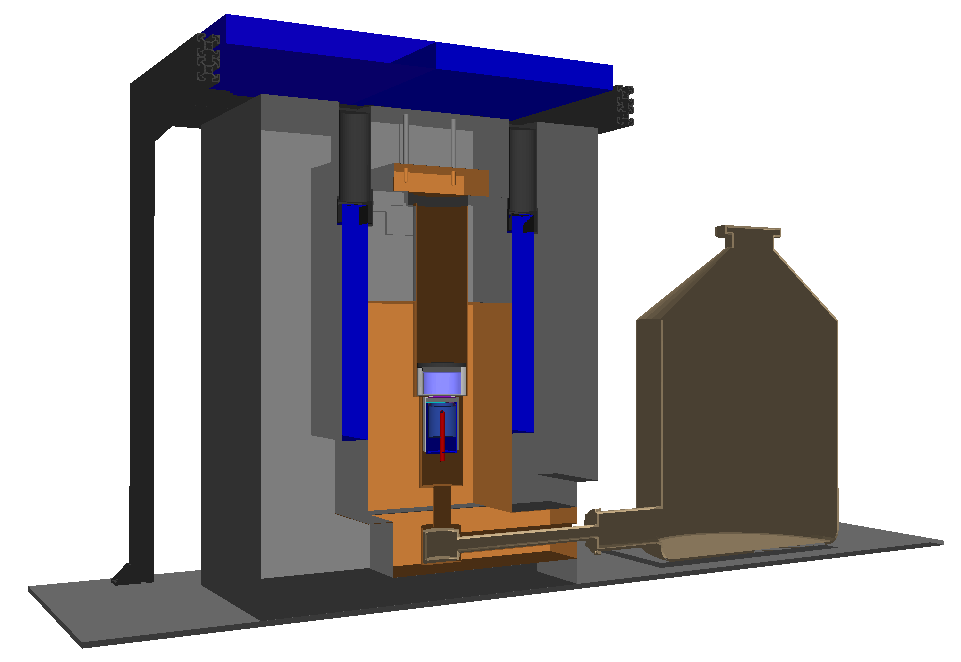}
			\caption{Monte Carlo model of the GeSparK detector with the new muon veto system based on internal plastic scintillators (blue).}
			\label{fig:GSMCModelNew}
		\end{center}
	\end{figure}
	The main difficulty in obtaining an almost unitary veto efficiency is to cover the lateral sides of the GeSparK detector with scintillator detectors.
	Since the lateral surface of the lower part of the lead shielding is too large to be covered by plastic scintillators, we have hypothesized to install them inside the passive shielding, between the copper and the lead shields.
    
    \begin{figure}[h]
        \centering
        \begin{subfigure}{0.45\textwidth}
            \centering
            \includegraphics[width=\textwidth]{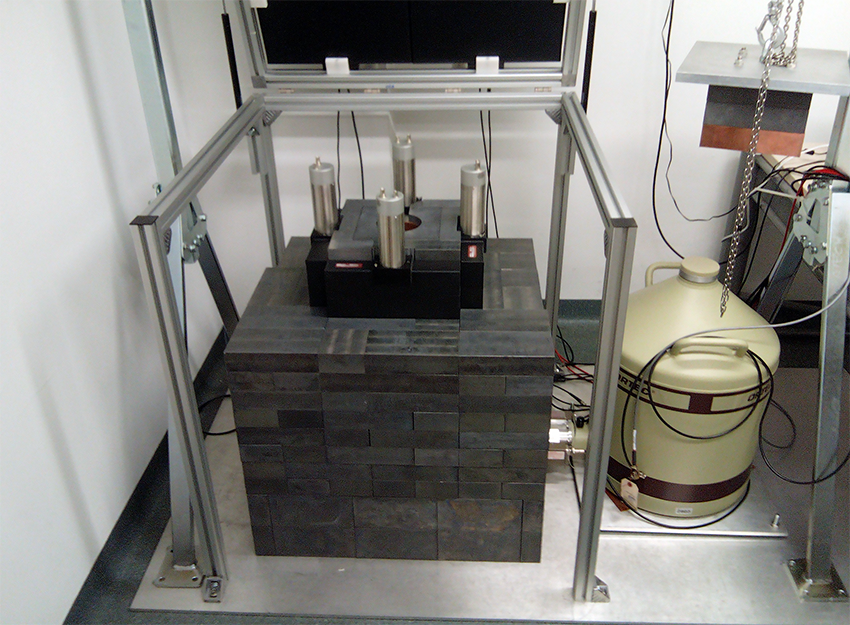}
            \caption{GeSparK detector during the installation of the internal plastic scintillators.}
            \label{fig:GeSparKNew1}
        \end{subfigure}
        \begin{subfigure}{0.45\textwidth}
            \centering
            \includegraphics[width=\textwidth]{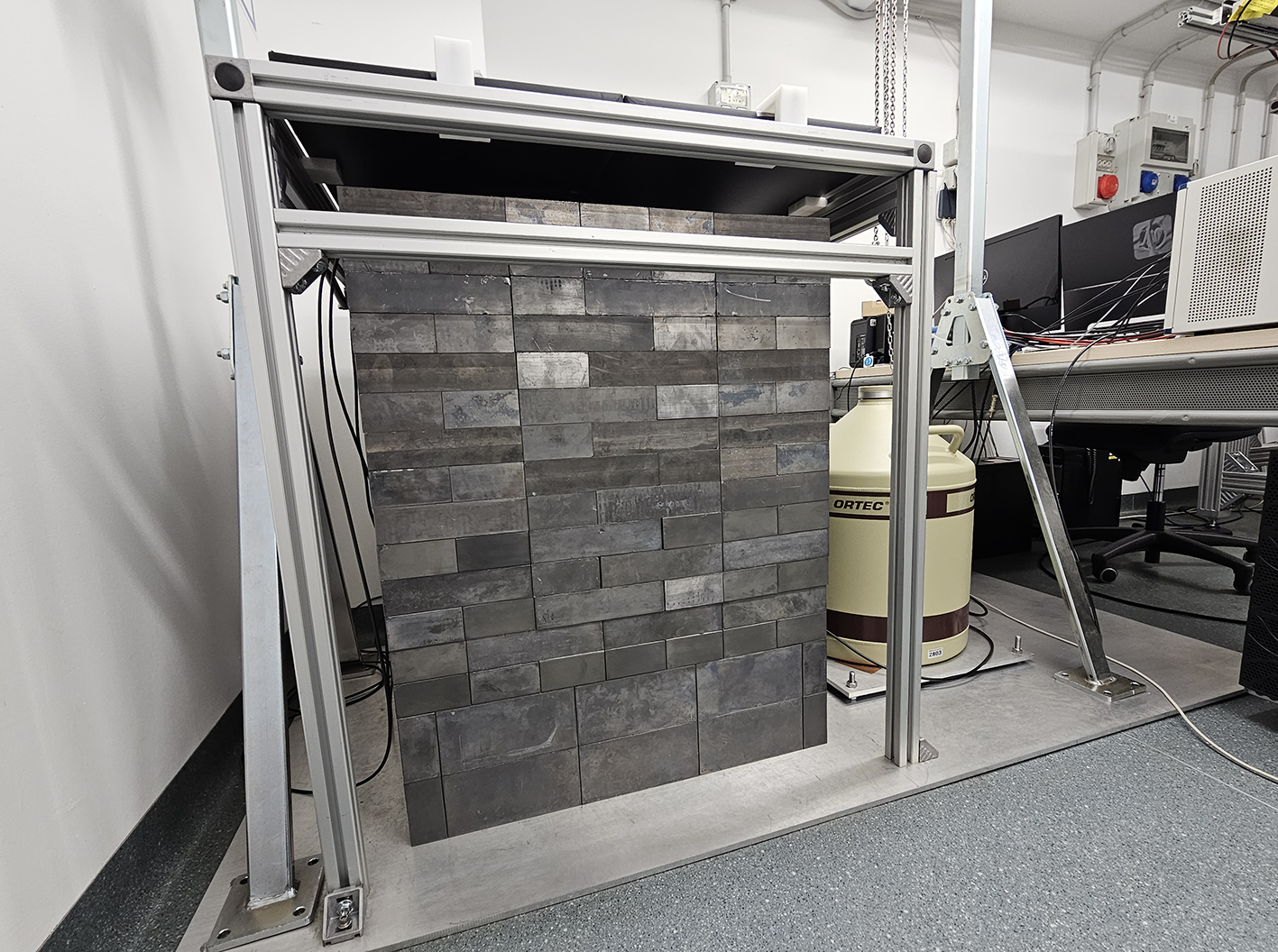}
            \caption{Final setup of the GeSparK detector.}
            \label{fig:GeSparKNew2}
        \end{subfigure}
        \caption{Pictures of the GeSparK detector with the new muon veto system based on internal plastic scintillators. Subfigure \subref{fig:GeSparKNew1} shows the GeSparK detector during the installation of the internal plastic scintillators and subfigure \subref{fig:GeSparKNew2} shows the final setup.}
        \label{fig:GeSparKNew}
    \end{figure}

    By exploiting the validated Monte Carlo simulation, the best configuration for the new veto system was determined. It is composed of six plastic scintillators in the following configuration: two on the top of the detector, as in the initial setup, and the other four around the LS and HPGe detector's copper shield, inside the lead shield, to cover the sides of the detectors, as shown in Figure \ref{fig:GSMCModelNew}. At the same time, we decided to improve the passive shielding by adding some layers of lead on the top.
	
	The Monte Carlo simulation of the new setup showed a potential veto efficiency of \qty{99.4(1)}{\%}. Based on these results we implemented this configuration.
	The resulting GeSparK system is shown in Figure \ref{fig:GeSparKNew}.

    In the next section, the measurement performed to verify the actual veto efficiency is discussed together with the updated sensitivities of the GeSparK detector.

	\section{Background measurements and detection limits}
	\label{S:4}
	
	\subsection{Background measurement and resulting veto efficiency}
	
	To measure the effective veto efficiency of the GeSparK system, a new background measurement was performed with no sample mixed in the liquid scintillator.
    \begin{table*}
		\centering
        \footnotesize
		\begin{tabular}{ccc}
			\toprule
            Measured (\qtyrange[range-units = single]{20}{3000}{keV}) & Measured (HPGe saturated) & Monte Carlo \\
			\midrule
			\qty{92.69(3)}{\%} & \qty{98.87(1)}{\%} & \qty{99.4(1)}{\%} \\
			\bottomrule
		\end{tabular}
		\caption{Measured background rejections in the whole spectrum from \qtyrange[range-units = single]{20}{3000}{keV} and considering only saturated events in HPGe detector (energy above \qty{3000}{keV)} together with the Monte Carlo veto efficiency estimation based on the final muon veto configuration. The reported errors include only the statistical contribution.}
		\label{tab:MuonVetoEfficiencyMCMISFinal}
	\end{table*}
 
	\begin{figure*}[ht]
		\begin{center}
			\includegraphics[width=0.9\textwidth]{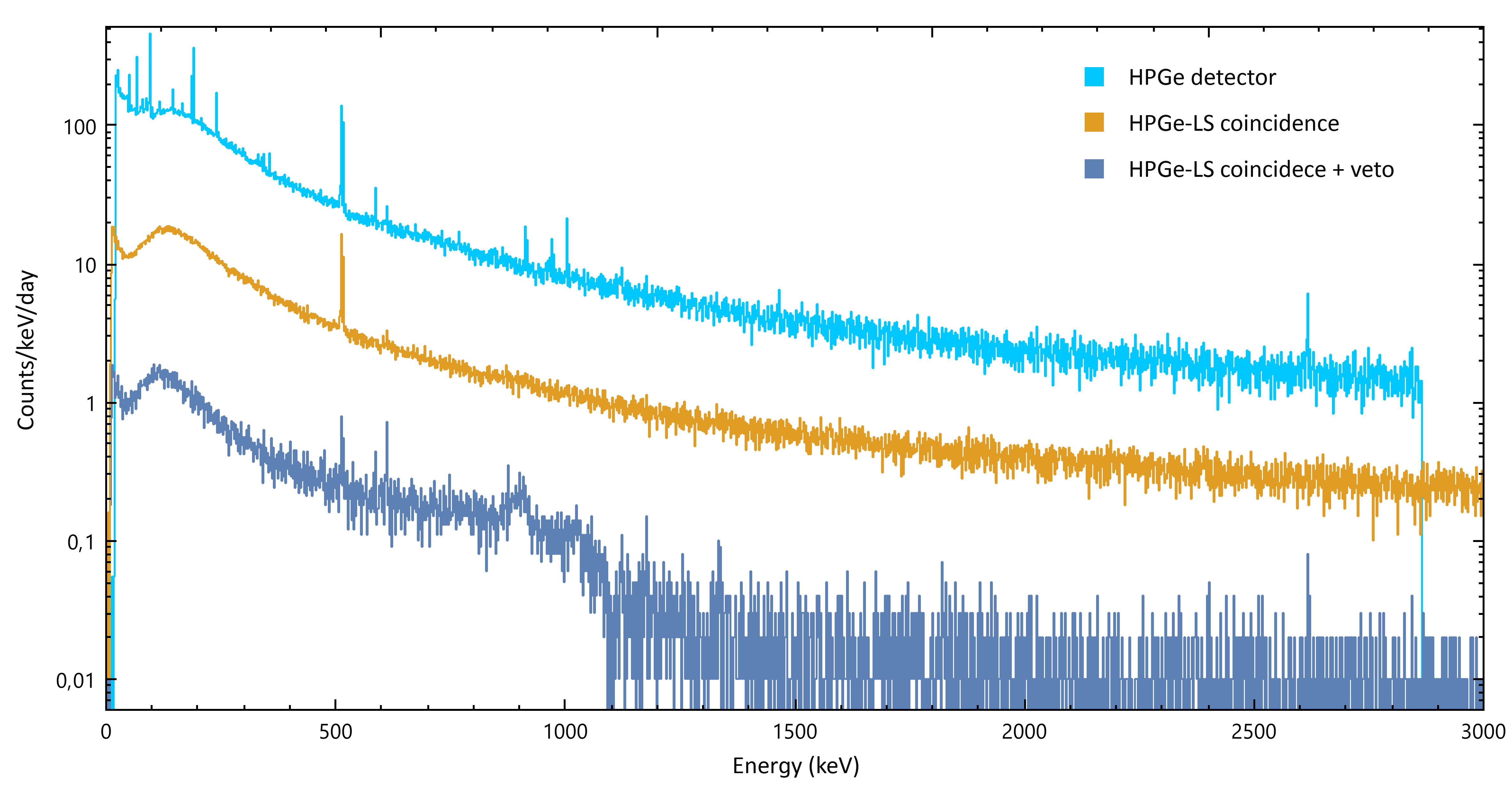}
			\caption{Germanium spectrum of background events in different configurations. From top to bottom: HPGe detector in single mode, HPGe in coincidence with LS detector, and HPGe in coincidence with LS detector and muon veto applied.}
			\label{fig:Gebackground}
		\end{center}
	\end{figure*}

    \begin{figure*}[ht]
		\begin{center}
			\includegraphics[width=0.9\textwidth]{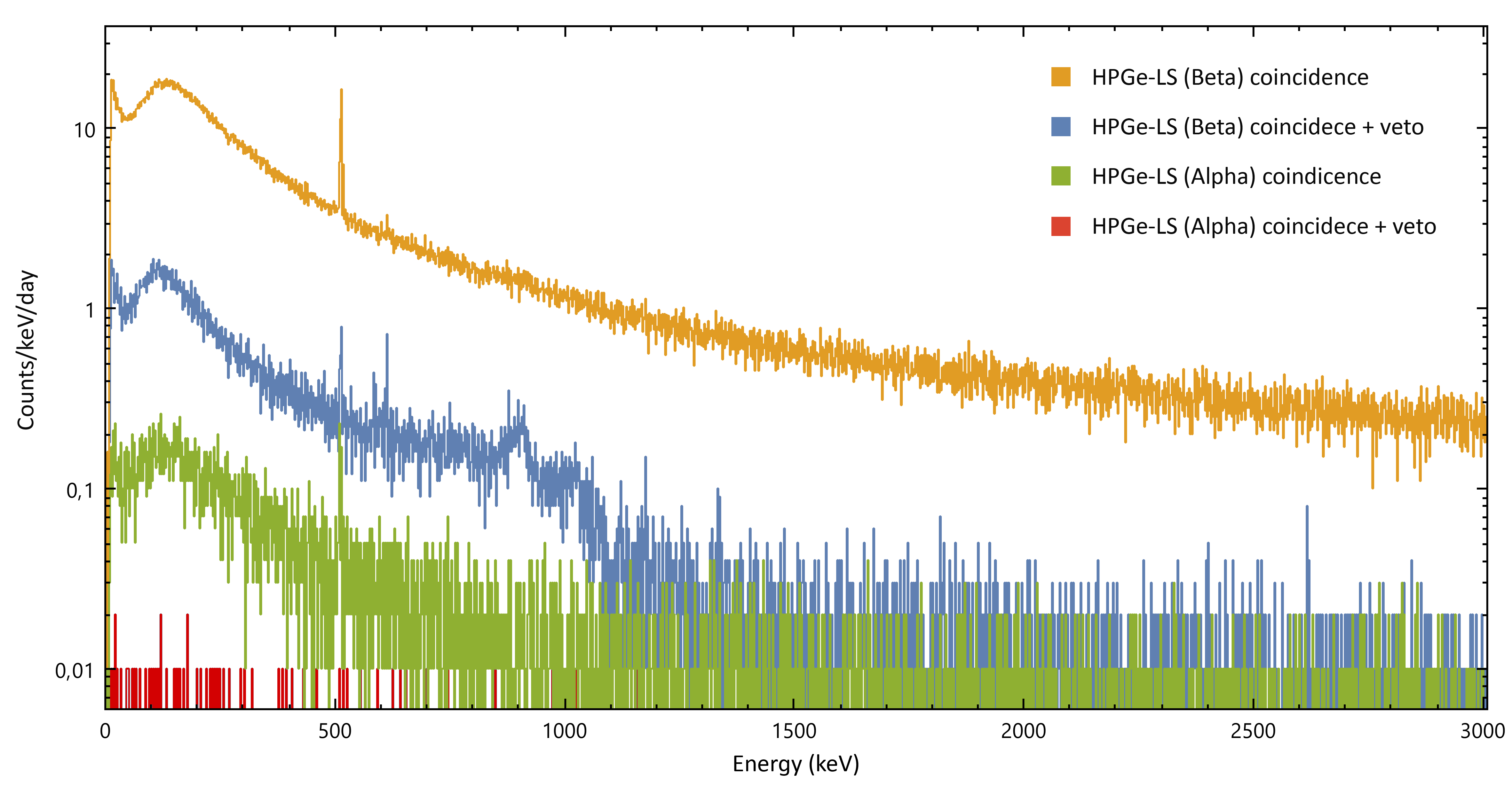}
			\caption{Germanium spectrum of background events in coincidence with LS detector with and without the muon veto applied and with and without $\alpha/\beta$ selection applied on LS detector events. From top to bottom: $\beta$ events without muon veto, $\beta$ events with muon veto, $\alpha$ events without muon veto, and $\alpha$ events with muon veto}
			\label{fig:GebackgroundAB}
		\end{center}
	\end{figure*}

    \begin{table*}[h!t]
		\centering
		\footnotesize
		\begin{tabular}{ccccccc}
			\toprule
			\multirow{3}{*}{Chain}&          &            & \multicolumn{2}{c}{HPGe-LS}            & \multicolumn{2}{c}{HPGe-LS-muon veto} \\ 
			                     & 	Nuclide	& $E_\gamma$ & \multicolumn{2}{c}{1.2 FWHM ROI rate}  & \multicolumn{2}{c}{peak* or 1.2 FWHM ROI rate} \\
			                     & 			& (keV)      & \multicolumn{2}{c}{($\mu$Hz)}          & \multicolumn{2}{c}{($\mu$Hz)}         \\
                                  &          &            & Old                 & New              & Old            & New                  \\
			\midrule
			$^{238}\textrm{U}$ 	& $^{238}\textrm{U} \; (\alpha)$ & 93 & $164\pm8$ & $338\pm6$ & $86\pm6$ & $35\pm2$ \\
			\midrule
			$^{235}\textrm{U}$ 	& $^{235}\textrm{U} \; (\alpha)$ & 186 &  $193\pm8$ & $295\pm6$ & $100\pm6$ & $19.5\pm1.5$ \\
			\midrule
			\multirow{3}{*}{$^{232}\textrm{Th}$} & $^{212}\textrm{Pb} \; (\beta )$ & 238.5 & $164\pm8$ & $213\pm5$ & $84\pm6$ & $17.7\pm1.4$ \\
			& $^{208}\textrm{Tl} \; (\beta)$ & 583 & $62\pm5$ & $72\pm3$ & $37\pm4$ & $6.7\pm0.9$* \\ 
			& $^{208}\textrm{Tl} \; (\beta)$ & 2615 & $9\pm2$ & $5.3\pm0.8$ & $5.2\pm1.4$ & $2.0\pm0.8$* \\
			\bottomrule
		\end{tabular}
		\caption{Comparison of the rates in the energy regions of the main gamma lines of the HPGe detector background spectrum when operated in single mode between previous and current configurations and considering the HPGe detector in coincidence with the liquid scintillation detector with and without the muon veto. The rates are evaluated in 1.2 FWHM ROI when no peaks are present in the spectra. For $^{208}\textrm{Tl}$ lines the rates with * are evaluated in the corresponding peaks.}
		\label{tab:bkgrate1}
	\end{table*}

    \begin{table*}[h!t]
		\centering
		\footnotesize
		\begin{tabular}{cccccc}
			\toprule
			{\multirow{2}{*}{ROI}}  & $E_\gamma$ &\multicolumn{2}{c}{HPGe-LS} & \multicolumn{2}{c}{HPGe-LS-muon veto}  \\ 
			             	   & (keV)      & \multicolumn{2}{c}{(mHz)}     & \multicolumn{2}{c}{(mHz)}         \\
                                      &            & Old      & New     & Old      & New                               \\
			\midrule
			$\textrm{e}^+/\textrm{e}^-$ peak & 511 & $0.389\pm0.012$ & $0.58\pm0.01$ & $0.214\pm0.009$ & $0.025\pm0.002$ \\ 
			\midrule
			Whole spectrum & 40-2700 & $47.63\pm0.13$ & $75.6\pm0.1$ & $24.44\pm0.10$ & $5.58\pm0.03$ \\ 
			\bottomrule
		\end{tabular}
		\caption{Comparison of the rates of the annihilation peak (\qty{511}{keV}) and the integral rate in the \qtyrange[range-units = single]{40}{2700}{keV} energy range in HPGe background spectra between previous and current configurations and considering the HPGe detector in coincidence with the liquid scintillation detector with and without the muon veto.}
		\label{tab:bkgrate2}
	\end{table*}

	\begin{table*}[h!t]
		\centering
		\footnotesize
		\begin{tabular}{ccccccccc}
			\toprule
			Chain & Nuclide (decay) & $E_\gamma$ (keV) & $\epsilon\cdot$BR (\%) & B & \multicolumn{2}{c}{MDA (mBq/kg)} & \multicolumn{2}{c}{MDC ($10^{-9}$g/g)}\\ 
                  &                 &                  &                        &   &   Old    &   New                &        Old     &     New              \\
			\midrule
			\multirow{3}{*}{$^{238}\textrm{U}$} & $^{226}\textrm{Ra} \; (\alpha )$ & 186 & 0.126 & 0 & 8.0 &  8.0 & 0.64 & 0.64\\
			& $^{214}\textrm{Pb} \; (\beta)$ & 352 & 0.804 & 21 & 26 & 11 & 2.1 & 0.89 \\
			& $^{214}\textrm{Bi} \; (\beta)$ & 609 & 0.618 & 23 & 27 & 15 & 2.2 & 1.2 \\ 
			\midrule
			\multirow{3}{*}{$^{235}\textrm{U}$} & $^{235}\textrm{U} \; (\alpha )$ & 143.7 & 0.503 & 0 & 9.7 & 2.0 & 0.12 & 0.025 \\
			& $^{211}\textrm{Bi} \; (\alpha)$ & 351 & 0.286 & 0 & 12 & 3.5 & 0.15 & 0.044 \\ 
			& $^{211}\textrm{Pb} \; (\beta)$ & 405 & 0.074 & 23 & 252 & 126 & 3.2 & 1.6 \\ 
			\midrule
			\multirow{3}{*}{$^{232}\textrm{Th}$} & $^{224}\textrm{Ra} \; (\alpha )$ & 241 & 0.108 & 0 & 9.4 & 9.4 & 2.3 & 2.3 \\
			& $^{228}\textrm{Ac} \; (\beta)$ & 911 & 0.290 & 14 & 49 & 26 & 12 & 6.3 \\ 
			& $^{208}\textrm{Tl} \; (\beta)$ & 583 & 0.429 & 21 & 43 & 21 & 11 & 5.4 \\ 
			\midrule
			& $^{60}\textrm{Co} \; (\beta)$ & 1173 & 0.843 & 2 & 14 & 4.0 & $3.3\cdot10^{-10}$ & $0.94\cdot10^{-10}$ \\ 
			& & 1332 & 0.757 & 3 & 15 & 5.4 & $3.6\cdot10^{-10}$ & $1.3\cdot10^{-10}$ \\ 
			\bottomrule
		\end{tabular}
		\caption{Comparison of the MDA and MDC of the detector between the previous and the current configuration for the main nuclides in $^{238}\textrm{U}$, $^{235}\textrm{U}$ and $^{232}\textrm{Th}$ chains and $^{60}\textrm{Co}$, all with a sample mass of \qty{100}{g} and a measurement time of \qty{31}{d}. The MDCs are relative to the parent nuclide of the chain assuming secular equilibrium.}
		\label{tab:MDA}
	\end{table*}

	By comparing the total event rate in the whole spectrum (\qtyrange[range-units = single]{20}{3000}{keV}) with and without the application of the muon veto (as applied in MC simulations), the resulting background reduction is \qty{92.69(3)}{\%}. This value is lower than the MC estimation and the main reason is due to the presence of other background sources that become more significant as the muon veto efficiency increases.
    To validate this hypothesis, the background reduction was evaluated on the saturated events in the HPGe detector (energy above \qty{3000}{keV}), which are almost completely due to muons events, and the resulting reduction is \qty{98.87(1)}{\%}, much more consistent with the simulation result. The residual difference is probably related to the geometrical model approximations and muon angular distribution differences. These results are summarised in Table \ref{tab:MuonVetoEfficiencyMCMISFinal}.
	In Figure \ref{fig:Gebackground}, it is possible to appreciate the reduction of the background of the HPGe acquired with and without the coincidence with the scintillator detector and the anti-coincidence with the cosmic veto system. By observing the spectrum shape it is evident that cosmic muon showers do not dominate the final spectrum and this explains the lower background rejection when compared to the MC one.

	The event rate of the HPGe detector, calculated in the energy interval \qtyrange[range-units = single]{40}{2700}{keV}), carried out by selecting only the events detected in coincidence with the LS detector, without the muon veto, is increased to \qty{75.6}{\mHz} with respect the previous configuration (\qty{47.6}{\mHz}). This is due to the lower threshold achievable by the LS detector thanks to the two-acquisition system (channel splitting), since the LS detector serves as the trigger for data acquisition, and the HPGe detector signal is recorded in coincidence. Nevertheless, after the application of the muon veto, the total event rate is reduced up to \qty{5.58}{\mHz}, a factor 4.4 lower than the previous veto configuration and the annihilation peak rate, only due to the cosmic showers, is reduced by a factor 8.6.
	
	The comparison of the event rate in the regions of interest of the gamma lines detected by the HPGe operated in single mode (light blue spectrum in Figure \ref{fig:Gebackground}) is performed between the HPGe operated in coincidence with the LS detector with and without the veto selection. The results are shown in Table \ref{tab:bkgrate1} for the gamma lines present in the HPGe detector background spectrum when operated in single mode and Table \ref{tab:bkgrate2} for the annihilation peak and integral rate.

    The LS detector of the GeSparK system can perform $\alpha/\beta$ discrimination by exploiting the different pulse shape of the signals produced by the two types of particles \cite{GS:2021}, reducing the measurement background. The HPGe spectrum of events acquired in coincidence to the LS detector by applying $\alpha/\beta$ discrimination is shown in Figure \ref{fig:GebackgroundAB}. Here it is possible to appreciate the background reduction obtained by selecting alpha events, in particular when combined with the muon veto (green and red spectra in Figure \ref{fig:GebackgroundAB}). In fact, muon coincidence helps removing those events misidentified as alpha-like.
	
	\subsection{Updated measurement sensitivity}
	
	The sensitivity of the GeSparK detector was updated according to the new background measurement in terms of minimum detectable activity (MDA) and minimum detectable concentration (MDC) for a sample mass of \qty{100}{g} measured for \qty{31}{d} (the same assumption of the previous paper). To this aim, we analyzed the HPGe detector background and used its statistical fluctuation in the region of interest (ROI) with a 95\% confidence level to estimate the MDA for a nuclide that has a gamma peak in that energy region.
	According to Currie \cite{Currie:1968}, the equation used to estimate the MDA is:
	\begin{equation}
	MDA\left(\frac{\textrm{Bq}}{\textrm{kg}}\right)=\frac{2.71+4.64\cdot\sqrt{\textit{B}}}{\epsilon\cdot BR\cdot m \cdot T}
	\end{equation}
	where B is the background integral area in the ROI, considering an energy interval corresponding to $1.2\cdot\text{FWHM}$, $\epsilon$ is detection efficiency, BR is the branching ratio of the observed decay, m is the mass of the sample (kg) and T is the measurement time (s). The MDC is computed from the MDA by converting the activity into the corresponding nuclide concentration, according to the following equation:
    \begin{equation}
	MDC\left(\frac{\textrm{g}}{\textrm{g}}\right)=MDA\left(\frac{\textrm{Bq}}{\textrm{kg}}\right)\frac{m_a}{\lambda*Na*1000}
	\end{equation}
    where $\lambda$ is the decay constant ($s^{-1}$), Na is the Avogadro number, and $m_a$ is the atomic mass (g/mol). 
	For nuclides in a chain, the concentration is referred to the progenitor of the chain assuming secular equilibrium.
	In Table \ref{tab:MDA}, we report the new MDA and MDC, compared with the previous one, for the main nuclides in the $^{238}\textrm{U}$ and $^{232}\textrm{Th}$ decay chains and $^{60}\textrm{Co}$.
    The number of events in the ROI was computed on the spectrum obtained by applying beta or alpha selection according to the decay mode of the considered nuclide.

	\section{Conclusions and future application}
	\label{S:5}
    To improve the sensitivities of the GeSparK detector we implemented a high efficiency veto system based on plastic scintillators surrounding the HPGe and LS detectors. This configuration allowed us to use smaller detectors than those required to cover the external surface of the lead shielding and to limit the total event rate on the plastic scintillators due to external radioactivity.
    The new veto system reached a background reduction of \qty{92.69(3)}{\%} in the energy region \qtyrange[range-units = single]{20}{3000}{keV}, a bit lower than the simulated one due to the presence of background sources not related to cosmic muons that become more significant as the veto efficiency increases.
    The high rejection capability allowed the GeSparK detector to reduce the background level and improve its sensitivities.
    In the final section of this work, we discussed the expected sensitivities on the main natural radioactivity sources, but the GeSparK detector is suitable also for Neutron Activation Analysis (NAA)\cite{NAA:2024} of liquid samples that can be dissolved in the LS, since the activation products usually decay beta minus, with a $\beta-\gamma$ signature. In this context, the new veto system of the GeSparK detector can improve the sensitivity on the relatively long-life activation products, such as $^{233}\text{Pa}$, daughter of the $^{233}\text{Th}$ and activation product of the $^{232}\text{Th}$, since in this cases the sensitivity is mainly limited by the detector residual background.
	

	
	\bibliographystyle{unsrt2authabbrvpp}
	\bibliography{Bibliography}

\begin{thebibliography}{1}

\bibitem{GS:2021}
G.~Baccolo et~al.
\newblock Development of a low background alpha–beta/gamma coincidence detector.
\newblock {\em Nucl Instrum Meth A}, 1003:165290, 2021.

\bibitem{Edler:2016}
R.~Edler.
\newblock {\em {Cocktails for liquid scintillation counting}}.
\newblock PerkinElmer LAS, Germany, 2016.

\bibitem{Agostinelli:2003}
S.~Agostinelli et~al.
\newblock {Geant4—a simulation toolkit}.
\newblock {\em Nucl Instrum Meth A}, 506(3):250 -- 303, 2003.

\bibitem{MuoniEne}
P.~Shukla and S.~Sankrith.
\newblock Energy and angular distributions of atmospheric muons at the earth.
\newblock {\em International Journal of Modern Physics A}, 33:1850175, 10 2018.

\bibitem{MuoniAng}
S.~Cecchini and M.~Spurio.
\newblock Atmospheric muons: experimental aspects.
\newblock {\em Geoscientific Instrumentation, Methods and Data Systems}, 1(2):185--196, 2012.

\bibitem{Currie:1968}
L.~A. Currie.
\newblock {Limits for qualitative detection and quantitative determination. Application to radiochemistry}.
\newblock {\em Anal Chem}, 40(3):586--593, 1968.

\bibitem{NAA:2024}
G.~Baccolo et~al.
\newblock {Radiopurity screening of materials for rare event searches by neutron activation at the TRIGA reactor of Pavia}.
\newblock {\em arXiv preprint arXiv:2411.02212}, 2024.

\end{thebibliography}

\end{document}